\documentclass[runinaddress,showpacs,twocolumn,prl]{revtex4}
\usepackage{amsmath}
\usepackage{amssymb}
\usepackage{graphicx}

\input{tcilatex}

\begin{document}

\title{Spin transverse force on spin current in an electric field}
\author{Shun-Qing Shen}
\affiliation{Department of Physics, and Center for Theoretical and Computational Physics,
The University of Hong Kong, Pokfulam Road, Hong Kong, China}
\date{Revised on July 27, 2005}

\begin{abstract}
As a relativistic quantum mechanical effect, it is shown that the electric
field exerting a transverse force on an electron spin 1/2 only if the
electron is moving. The spin force, analogue to the Lorentz for on an
electron charge in a magnetic field, is perpendicular to the electric field
and the spin current which spin polarization is projected anong the electric
field. This spin-dependent force can be used to understand the
zitterbewegung of electron wave packet with spin-orbit coupling and is
relevant to the generation of the charge Hall effect driven by the spin
current in semiconductors.
\end{abstract}

\pacs{85.75.-d, 72.20.My, 71.10.Ca}
\maketitle

Recent years spintronics has became an emerging field because of its
potential application to semiconductor industry, and more and more
attentions are focused on the effect of spin-orbit coupling in metals and
semiconductors.\cite{Prinz98Science} The spin-orbit coupling is a
relativistic effect describing the interaction of the electron spin,
momentum and electric field, and provides a route to manipulate and to
control the quantum spin states via external fields.\cite{Datta90,Dyakonov71}
It is desirable to understand the motion of electron spin with spin-orbit
coupling in an electromagnetic field. In electrodynamics it is known that a
magnetic field would exert a Lorentz force on an electric charge if it is
moving. This Lorentz force can generate a lot of fundamental phenomena such
as the Hall effect in solid.\cite{Hall} The interaction of the spin in the
electromagnetic field behaves as if the spin is a gauge charge and the
interaction is due to the SU(2) gauge field.\cite{Anandan89} It is essential
that the electron spin is an intrinsic quantum variable, not a just
classical tiny magnetic moment. The physical meaning of the interaction is
closely associated to the Aharonov-Casher effect and the Berry phase.

In this paper it is found that an electric field exerts a transverse force
on an electron spin if it is moving and the spin is projected along the
electric field. The force is proportional to the square of electric field,
and the spin current projected along the field, and its direction is always
perpendicular to the electric field and the speed of the spin. The force
stems from the spin-orbital coupling which can be derived from the Dirac
equation for an electron in a potential in the non-relativistic limit or the
Kane model with the $\mathbf{k}\cdot \mathbf{p}$ coupling between the
conduction band and valence band. From an exact solution of a single
electron with the spin-orbit coupling it is heuristic to understand that the
zitterbewegung of electronic wave packet is driven by the spin transverse
force on a moving spin. The role of spin transverse force is also discussed
in the spin Hall effect and its reciprocal effect driven by the pure spin
current in semiconductors.

We start with the Dirac equation of an electron in a confining potential $V$
and a vector potential $\mathbf{A}$ for a magnetic field, $\mathbf{B}=\nabla
\times \mathbf{A}$, 
\begin{equation}
i\hbar \frac{\partial }{\partial t}\Psi =\left[ c\alpha \cdot \left( \mathbf{%
p}+\frac{e}{c}\mathbf{A}\right) +mc^{2}\gamma +V\right] \Psi ,
\end{equation}%
where $\alpha $ and $\gamma $ are the $4\times 4$ Dirac matrices. $m$ and $e$
are the electron mass and charge, respectively, and $c$ is the speed of
light. We let $\Psi =\left( 
\begin{array}{c}
\varphi \\ 
\chi%
\end{array}%
\right) e^{-imc^{2}t/\hbar }$ such that the rest mass energy of electron is
removed from the energy eigenvalue of the electron. In the nonrelativistic
limit, $\chi $ is a very small component, $\chi \approx \frac{1}{mc^{2}-V}%
c\sigma \cdot \left( \mathbf{p}+\frac{e}{c}\mathbf{A}\right) \varphi $ where 
$\sigma $ are the Pauli matrices. Thus the component $\varphi $ of the
wavefunction satisfies the following equation, $i\hbar \frac{\partial }{%
\partial t}\varphi \approx H\varphi $, where%
\begin{equation}
H\approx \frac{\left( \mathbf{p}+\frac{e}{c}\mathbf{A}\right) ^{2}}{2m}%
+V_{eff}+\mu _{B}\sigma \cdot B+\frac{\hbar \left( \mathbf{p}+\frac{e}{c}%
\mathbf{A}\right) }{4m^{2}c^{2}}\cdot \left( \sigma \times \nabla V\right) 
\text{,}  \label{soc}
\end{equation}%
where $\mu _{B}=e\hbar /2mc$ and $V_{eff}=V+\frac{\hbar ^{2}}{8m^{2}c^{2}}%
\nabla ^{2}V.$ In the last step we neglect the higher order terms of
expansion. Thus the Dirac equation is reduced to the Schr\"{o}dinger
equation with the spin-orbit coupling. The same form of effective spin-orbit
coupling and Zeeman splitting can be also derived from the $8\times 8$ Kane
model that takes into account only the $\mathbf{k\cdot p}$ coupling between
the $\Gamma _{6}^{c}$ conduction band and the $\Gamma _{8}^{v}$ and $\Gamma
_{7}^{v}$\ valence bands, although the effective mass, the effective Lande
g-factor, and the effective coupling coefficients have to be introduced as
material-specific parameters such as $m\rightarrow m^{\ast }$, $\mu
_{B}\rightarrow g\mu _{B}/2$, and $\hbar ^{2}/(4m^{2}c^{2})\rightarrow
r_{41}^{6c6c}$. \cite{Winkler03}

In the Heisenberg picture the kinetic velocity is 
\begin{equation}
\mathbf{v}=\frac{1}{i\hbar }\left[ \mathbf{r},H\right] =\frac{1}{m}\left[ 
\mathbf{p}+\frac{e}{c}\left( \mathbf{A}+\mathcal{A}\right) \right]
\end{equation}%
where $\mathcal{A=}\frac{\hbar }{4mce}\sigma \times \nabla V$ and comes from
the spin-orbit coupling. It indicates clearly that $\mathcal{A}$ plays a
role of a SU(2) gauge vector potential. The spin dependence of the gauge
field can separates the charged carriers with different spins in cyclotron
motion experimentally.\cite{Rokhinson04prl} Even though we have $[p_{\alpha
},p_{\beta }]=0$ for canonical momentum, the analogous commutators do not
vanish for the kinetic velocity%
\begin{equation}
\lbrack v_{\alpha },v_{\beta }]=-i\frac{\hbar e}{m^{2}c}\epsilon _{\alpha
\beta \gamma }\mathcal{B}_{\gamma }+\frac{e^{2}}{m^{2}c^{2}}[\mathcal{A}%
_{\alpha },\mathcal{A}_{\beta }]
\end{equation}%
where the total magnetic field $\mathcal{B}=\mathbf{B}+\nabla \times 
\mathcal{A},$ and $\nabla \times \mathcal{A}=\hbar \left[ \sigma \cdot
(\nabla ^{2}V)-(\sigma \cdot \nabla )\nabla V\right] /\left( 4mce\right) .$
Notice that $[\mathcal{A}_{\alpha },\mathcal{A}_{\beta }]=2\hbar
^{2}i(\sigma \cdot \nabla V)\epsilon _{\alpha \beta \gamma }\partial
_{\gamma }V/(4mce)^{2}.$ We can derive the quantum mechanical version of the
force,

\begin{equation}
m\frac{d\mathbf{v}}{dt}=F_{h}+F_{g}+F_{f}  \label{force}
\end{equation}%
with 
\begin{subequations}
\begin{eqnarray}
F_{h} &=&-\frac{e}{c}\mathbf{v}\times \mathcal{B}-\nabla \left( V_{eff}+\mu
_{B}\sigma \cdot B\right) \\
F_{g} &=&\frac{\mu _{B}}{2mc^{2}}[\sigma (\mathbf{B}\cdot \nabla V)-\mathbf{B%
}(\sigma \cdot \nabla V)], \\
F_{f} &=&\frac{\hbar }{8m^{2}c^{4}}(\sigma \cdot \nabla V)(\mathbf{v}\times
\nabla V)
\end{eqnarray}%
This is the quantum mechanical analogue of Newton's second law. Of course we
should notice that this is just an operator equation. The uncertainty
relationship tells that the position and momentum cannot be measured
simultaneously, and there is no concept of force in quantum mechanics. To
see the physical meaning of the equation, we take the expectation values of
both sides with respect to a Heisenberg state $\left\vert \Phi \right\rangle 
$ which does not varies with time. The expectation values of the observable
describe the motion of the center of the wave package of electrons. In this
sense we have an equation of the force experienced by the moving electron.
Actually the first term $F_{h}$ in Eq. (\ref{force}) is the Lorentz force
for a charged particle in a magnetic field $\left\langle \mathcal{B}%
\right\rangle $ which contains the contribution from the SU(2) gauge field $%
\mathcal{A}$ as well as the conventional electromagnetic field. We have
recovered the Ehrenfest theorem as one of the examples of the corresponding
principle in quantum mechanics. The term, $\left\langle \nabla (\mu
_{B}\sigma \cdot B)\right\rangle ,$ results from the non-uniform magnetic
field. Its role was first realized in the Stern-Gerlach experiment, where a
shaped magnet is used to generate a non-uniform magnetic field to split the
beam of silver atoms. In the classical limit it is written as the
interaction between the magnetic momentum $\mu =-\left\langle \mu _{B}\sigma
\right\rangle $ and magnetic field. This spin force depends on the spin.
Recently it is proposed that the force can generate a pure spin current if
we assume $\nabla B_{z}$ is a constant.\cite{Zhang04xxx} In the term $F_{g}$
we can also use $\mu $ to replace the spin. It is non-zero only when the
electric and magnetic fields coexist, as suggested by Anandan and others.%
\cite{Anandan89} This term will play an essential role in generating spin
Hall current in two-dimensional Rashba system, which we will discuss it
later.

The last term, $F_{f}$, comes from the SU(2) gauge potential or spin-orbital
coupling. As the force is related to the Planck constant it has no
counterpart in classic mechanics and is purely quantum mechanic effect. The
force is irrelevant of the magnetic field. In the classical limit we cannot
simply use the magnetic momentum $\mu $ to replace the spin $\sigma $ in the
potential $\mathcal{A}$. Otherwise the force vanishes. To see the physical
meaning of the force, we write the spin force for a single electron on a
quantum state in a compact form,

\end{subequations}
\begin{equation}
\left\langle F_{f}\right\rangle =\frac{e^{2}\left\vert \mathcal{E}%
\right\vert }{4m^{2}c^{4}}\mathbf{j}_{s}^{\mathcal{E}}\times \mathcal{E}
\label{spinforce}
\end{equation}%
where the spin current is defined conventionally, $\mathbf{j}_{s}^{\mathcal{E%
}}=\frac{\hbar }{4}\left\langle \{\mathbf{v},\sigma \cdot \mathcal{E}%
/\left\vert \mathcal{E}\right\vert \}\right\rangle ,$\ and in the last step
the relation $\{\mathcal{A},\sigma \cdot \mathcal{E}\}=0$ has been used.
This is the main result in this paper. The force is proportional to the
square of electric field $\mathcal{E}$ and the spin current which
polarization is projected along the field. It is important to note that an
electron in a spin state perpendicular to the electric field will not
experience any force. Comparing with a charged particle in a magnetic field, 
$\mathbf{j}_{c}\times \mathbf{B,}$ where $\mathbf{j}_{c}$ is a charge
current density, the spin force is nonlinear to the electric field and
depends on the spin state of electron.

We discuss several examples relevant to the spin force. Though the gauge
field $\mathcal{A}$ provides a spin-dependent magnetic field $\nabla \times 
\mathcal{A}$, the Lorentz force caused by the field on the charge will
vanish in a uniform electric field $\nabla V=e\mathcal{E}.$ The spin depend
force $F_{g}$ also vanishes in the absence of magnetic field. Here we
consider the motion of an electron confining in a two-dimensional plane
subjected to a perpendicular electric field, 
\begin{equation}
H=\frac{\mathbf{p}^{2}}{2m}+\lambda (\mathbf{p}_{x}\sigma _{y}-\mathbf{p}%
_{y}\sigma _{x})  \label{ham}
\end{equation}%
\ where $\lambda =\hbar e\mathcal{E}/(4m^{2}c^{2})$ from Eq.(\ref{soc}).
This can be regarded as counterpart of a charged particle in a magnetic
field. On the other hand it has the same form of the Rashba coupling in a
semiconductor heterojuction with the structural inversion asymmetry,\cite%
{Rashba60} where the spin-orbit coupling is induced by the offsets of
valence bands at the interfaces and the structure inversion asymmetry.\cite%
{Winkler03} A typical value of this coefficient $\lambda $ is of order $%
10^{-4}c$ ($c$ the speed of light), and can be adjusted by an external field.%
\cite{Nitta97apl} Because of the spin-orbit coupling the electron spin will
precess with time,%
\begin{equation}
\frac{d\sigma (t)}{dt}=\frac{2\lambda }{\hbar }\sigma (t)\times \left( 
\mathbf{p}\times \hat{z}\right) .
\end{equation}%
Since the momentum $\mathbf{p}$ is a good quantum number, without loss of
generality, we take $\mathit{p=p}_{x},$ just along the x direction.
Correspondingly the wave function in the position space $\left\langle r|\Phi
\right\rangle =\exp (ip_{x}x)\chi _{s}/\sqrt{L}$ where $\chi _{s}$ is the
initial spin state. Equivalently the spin-orbit coupling provides an
effective magnetic field along the $y$ direction, $\mathbf{B}_{eff}=\lambda 
\mathbf{p}_{x}\hat{y}$. This problem can be solved analytically, and the
electron spin precesses in the spin x-z plane,\cite{Shen04prb} $\sigma
_{x}(t)=\sigma _{x}\cos \omega _{c}t-\sigma _{z}\sin \omega _{c}t,$ $\sigma
_{z}(t)=\sigma _{z}\cos \omega _{c}t+\sigma _{x}\sin \omega _{c}t,$ and $%
\sigma _{y}(t)=\sigma _{y}$ where the Larmor frequency $\omega
_{c}=2p_{x}\lambda /\hbar .$ The spin $\sigma _{z}(t)$ varies with time and
the spin current is always along the x direction, $\left\langle \mathbf{j}%
_{s}^{z}\right\rangle =\frac{\hbar }{2}\frac{p_{x}}{m}\left( \left\langle
\sigma _{z}\right\rangle \cos \omega _{c}t+\left\langle \sigma
_{x}\right\rangle \sin \omega _{c}t\right) \hat{x}$ where $\left\langle
...\right\rangle $ means the expectation value over an initial state $%
\left\langle r|\Phi \right\rangle $. As a result the spin transverse force
on the spin is always perpendicular to the x direction. Correspondingly the
kinetic velocity$v_{x}$ and $v_{y}$ at a time $t$ are 
\begin{subequations}
\begin{eqnarray}
\left\langle v_{x}\right\rangle _{t} &=&\frac{p_{x}}{m}+\lambda \left\langle
\sigma _{y}\right\rangle ; \\
\left\langle v_{y}\right\rangle _{t} &=&-\lambda \left( \left\langle \sigma
_{x}\right\rangle \cos \omega _{c}t-\left\langle \sigma _{z}\right\rangle
\sin \omega _{c}t\right) ,
\end{eqnarray}%
respectively. Though $p_{y}=0$ the kinetic velocity $\left\langle
v_{y}\right\rangle _{t}$ oscillates with the frequency $\omega _{c}$ while $%
\left\langle v_{x}\right\rangle _{t}$ remains constant. The y-component of
the position is 
\end{subequations}
\begin{equation}
\left\langle y\right\rangle _{t}=\left\langle y\right\rangle _{t=0}-\frac{%
\hbar }{p_{x}}\sin \frac{\omega _{c}t}{2}\left( \left\langle \sigma
_{x}\right\rangle \cos \frac{\omega _{c}t}{2}-\left\langle \sigma
_{z}\right\rangle \sin \frac{\omega _{c}t}{2}\right) .
\end{equation}%
If the initial state is polarized along y direction the electron spin does
not vary with time as it is an energy eigen state of the system, as
discussed by Datta and Das.\cite{Datta90} In this case the spin current $%
\left\langle \mathbf{j}_{s}^{z}\right\rangle $ carried by the electron is
always zero and the spin transverse force is zero. Thus $\left\langle
v_{y}\right\rangle _{t}=0.$ If the initial state is along the spin z
direction at $t=0$, i.e., $\left\langle \sigma _{z}\right\rangle =s=\pm 1,$
it is found that $\left\langle v_{y}\right\rangle _{t,s}=-s\lambda \sin
\omega _{c}t.$ Different spins will move in opposite directions. It can be
understood that the spin precession makes the spin current which
polarization is projected along the electric field changes with time such
that the spin force along the y direction also oscillates with the frequency 
$\omega _{c}$. This force will generate a non-zero velocity of electron
oscillating along the y direction. Though $\left\langle v_{y}\right\rangle
_{t,s=1}-\left\langle v_{y}\right\rangle _{t,s=-1}=-2\lambda \sin \omega
_{c}t,$ the velocity does not contribute to the spin current along the y
direction, i.e. $\left\langle \left\{ v_{y},\sigma _{z}\right\}
\right\rangle _{t,s}=0.$ The trajectory oscillates with the time. The
amplitude of the oscillation is $\hbar /p_{x}$ and the frequency is $\omega
_{c}=2p_{x}\lambda /\hbar .$ For a typical two dimensional electron gas the
electron density is $n_{e}=10^{11}\sim 10^{12}/$cm$^{2}$ and the wave length
near the Fermi surface $\hbar /p_{x}\sim 3-10$nm. For a typical Rashba
coupling $\lambda =10^{-4}c$, $\omega _{c}=0.3\sim 1.0\times 10^{-14}s.$ The
rapid oscillation of the electron wave packet is known in literatures as the
zitterbewegung of electron as a relativistic effect, which is usually
regarded as a result of admixture of the positron state in electron wave
packet as a relativistic effect.\cite{Schrodinger30} In semiconductors the
Rashba coupling reflects the admixture of the particle and hole states in
the conduction and valence bands. Recently Schliemann et al obtained the
solution of the trajectory and proposed that this effect can be observed in
a III-V zinc-blende semiconductor quantum wells.\cite{Schliemann05prl} In
the p-doped semiconductors described by the Luttinger model there exists
also a spin force, and will generate the zitterbewegung as calculated by
Jiang et al.\cite{Jiang04xxx} Though the spin transverse force on a moving
spin is very analogous to the Lorentz force on a moving charge, because of
spin precession, its effect is completely different with the motion of a
charged particle in a magnetic field, where the amplitude of the Lorentz
force is constant and the charged particle will move in a circle. The
zitterbewegung of the electronic wave package near the boundary will cause
some edge effect as shown in recent numerical calculations.\cite{edge} The
edge effect is determined by the electron momentum. The smaller the
momentum, the larger the edge effect. The amplitude and frequency of the
zitterbewegung satisfy a relation that $\left( \hbar /2p_{x}\right)
(2p_{x}\lambda /\hbar )=\lambda $, which is the amplitude of oscillation of
the velocity $v_{y}$. In Fig.1 it is illustrated that for two electrons with
different spins experience opposite forces in an electric field.

\begin{figure}[tbp]
\includegraphics*[width=8.5cm]{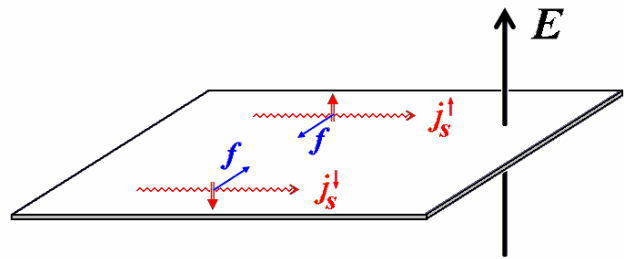}
\caption{ The electric field exerts opposite forces on electrons with
different spins polarized along the field. }
\end{figure}

Furthermore we consider a two-dimensional electron gas lacking both the bulk
and structural inversion symmetries. A Dresselhaus term $\beta (\mathbf{p}%
_{x}\sigma _{x}-\mathbf{p}_{y}\sigma _{y})$ will be included in the total
Hamiltonian in Eq. (\ref{ham}).\cite{Dresselhaus55} In this model the spin
force formula gives\cite{Li05prb}

\begin{equation}
\left\langle F_{f}\right\rangle =\frac{4m^{2}}{\hbar ^{2}}(\lambda
^{2}-\beta ^{2})(j_{s}^{z}\times \hat{z})
\end{equation}%
for each moving electron. First of all the force disappears at the symmetric
point of $\lambda =\pm \beta .$ At this point the operator $\sigma _{x}\pm
\sigma _{y}$ is a good quantum number an there is no spin flip in the
system. For $\lambda \neq \pm \beta ,$ the moving electron will experience a
spin-dependent force and the force will change its sign near $\lambda =\beta 
$. A heuristic picture from this formula is that when a non-zero spin
current $j_{s}^{z}$ goes through this system the spin-orbit coupling exerts
the spin transverse force on the spin current, and drives electrons to form
a charge Hall current perpendicular to the spin current. The\ injected spin
current can be generated in various ways, such as by the spin force $\nabla (%
\mathbf{\mu }\cdot \mathbf{B)}$ \ \cite{Zhang04xxx} and circularly or
linearly polarized light injection.\cite{Sipe,Hankiewicz05xxx} For instance
we assume the spin current $j_{s}^{z}$ be generated by the linear polarized
light injection which is proportional to\ the transition rate of from the
valence band to the conduction band with finite momentum and the life time
of electrons at the excited states. In the relaxation time $\tau $
approximation in a steady state the drift velocity orthogonal to the spin
current is $\left\langle v_{y}\right\rangle =$ $\frac{4m}{\hbar ^{2}}%
(\lambda ^{2}-\beta ^{2})j_{s}^{z}\tau $ if $\tau $ is not so long, i.e. \ $%
2p_{x}\sqrt{\lambda ^{2}+\beta ^{2}}\tau <<\hbar .$ This non-zero drift
velocity will form a Hall current orthogonal to the the spin current. This
is the charge Hall effect driven by the spin current. In ferromagnetic metal
or diluted magnetic semiconductors the charge current is spin polarized it
can generate the spin polarized Hall current via the spin transverse force.
Thus the spin transverse force can be also regarded a driven force of an
anomalous Hall effect\cite{AHE} and the spin-resolved Hall effect.\cite%
{Bulgakov99prl,Li05prb} A detailed calculation for this Hall conductance is
given by the Kubo formula as a linear response to the field $B=\Delta Bx\hat{%
z}$. This field will generate a spin force, $-\nabla (g\mu _{B}B\cdot \sigma
)=-g\mu _{B}\Delta B\sigma _{z}\hat{x},$ which will circulate a spin current
along the x direction, and furthermore the spin-orbit coupling provides a
driving force to generate a transverse charge current, $j_{c,y}$. In the
clear limit the Hall conductance $\sigma _{xy}=j_{c,y}/(g\mu _{B}\Delta B)=0$
for $\lambda =\pm \beta $ and $(e/4\pi \hbar )(\lambda ^{2}-\beta
^{2})/\left\vert \lambda ^{2}-\beta ^{2}\right\vert $.\ However following
Inoue et al and Mishchenko et al\cite{Inoue05prb} the inclusion of
impurities scattering will suppress the Hall conductance completely just
like the spin Hall effect. On the other hand numerical calculation in
mesoscopic systems shows the existence of the effect.\cite{Hankiewicz05xxx}
Another example is the two-dimensional p-doped system with the cubic Rashba
coupling,\cite{Schliemann05prb} 
\begin{equation}
H=\frac{\mathbf{p}^{2}}{2m}+i\alpha (\mathbf{p}_{+}^{3}\sigma _{-}-\mathbf{p}%
_{-}^{3}\sigma _{+})
\end{equation}%
where $\sigma _{\pm }$ are spin increasing and decreasing operators, and $%
\mathbf{p}_{\pm }=\mathbf{p}_{x}\pm i\mathbf{p}_{y}$. The spin force on the
moving electron in this system is $F_{f}=(2m\mathbf{p}\alpha /\hbar )^{2}(%
\mathbf{j}_{s}^{z}\times \hat{z}).$ The linear response theory gives the
Hall conductance $-9e/(2\pi \hbar )$ which is robust against the vertex
correction from impurities scattering. Calculations by means of the
Green-Keldysh function technique and linear response theory \cite%
{Zhang04xxx,Hankiewicz05xxx} show the existence of charge Hall effect driven
by the spin current, and the Onsager relation between the charge Hall effect
and its reciprocal. The key features of the numerical results are in good
agreement with the picture of spin force qualitatively.

In conclusion, an electric field exerts a transverse force on a moving spin
just like a magnetic field exerts a Lorentz force on a moving charge. This
force is proportional to the square of electric field and the spin current
with spin projected along the field. This is a purely relativistic quantum
mechanical effect. As the origin of the force the spin current should be
also observable physically. From the solution of the motion of a single
electron in an electric field, the zitterbewegung of electronic wave packet
in the spin-orbit coupling can be regarded as an explicit consequence of
this force. Due to the similarity of this spin transverse force and the
Lorentz force, the spin transverse force plays a similar role in the
formation of the charge Hall effect driven by the spin current and the spin
Hall effect driven by the charge current as the Lorentz force does in the
Hall effect in a magnetic field.

The author thanks F. C. Zhang for helpful discussion. This work was
supported by the Research Grant Council of Hong Kong (No.: HKU 7039/05P),
and a CRCG grant of The University of Hong Kong.

\end{document}